%% file: bibliography.tex
\documentclass[aps,prl,twocolumn,showpacs,superscriptaddress,groupedaddress]{revtex4}  
\usepackage{graphicx}  
\usepackage{dcolumn}   
\usepackage{bm}        
\usepackage{amssymb}   
\usepackage{amsmath}   
\usepackage{algorithm}
\usepackage{algorithmic}

\begin{document}


\title{Retrieving fields from proton radiography without source profiles}
\input authorlist.tex        
\date{\today}

\begin{abstract}
Proton radiography is a technique in high energy density science to diagnose magnetic and/or electric fields in a plasma by firing a proton beam and detecting its modulated intensity profile on a screen. Current approaches to retrieve the integrated field from the modulated intensity profile require the unmodulated beam intensity profile before the interaction, which is rarely available experimentally due to shot-to-shot variability. In this paper, we present a statistical method to retrieve the integrated field without needing to know the exact source profile. We apply our method to experimental data, showing the robustness of our approach. Our proposed technique allows not only for the retrieval of the path-integrated fields, but also of the statistical properties of the fields.
\end{abstract}

\maketitle

\section{\label{sec:introduction}Introduction}

Proton radiography is a popular method to diagnose magnetic and/or electric fields in laser-produced plasmas and inertial confinement fusion experiments \cite{nilson2006magnetic, tzeferacos-18, willingale2011high, romagnani2010observation, li2007observation, huntington2015observation, borghesi2002macroscopic}. In proton radiography, a low density proton beam is fired into a plasma, gets deflected by its electric and magnetic fields, and the beam intensity profile is measured on a screen placed at some distance. The modulated intensity of the beam is related to the deflection of the proton beam and thus related to the integrated magnetic and electric field. Therefore, by comparing the unmodulated beam intensity and the modulated intensity one can retrieve the integrated magnetic and electric fields.


Earlier attempts to reconstruct fields from proton radiography utilized Poisson's equation solver \cite{kugland-12-prorad} and a diffusion model \cite{graziani2017inferring}. However, those approaches use linear approximations of the proton radiography forward model, while most of the interesting interactions happen in the non-linear regime.
Later, Graziani \textit{et al.} \cite{graziani2017inferring} and Kasim \textit{et al.} \cite{kasim-17-prorad} independently realized that integrated field retrieval with proton radiography is a subclass of the optimal transport problem first posed in 1781 \cite{monge-1781}. Establishing the connection between proton radiography and optimal transport opens up numerous algorithms for proton radiography field reconstruction even in the non-linear regime. This has been shown by Kasim \textit{et al.} \cite{kasim-17-prorad} and Bott \textit{et al.} \cite{bott-17-prorad} by applying off-the-shelf optimal transport algorithms \cite{aurenhammer-1998-minkowski,sulman-11-efficient} to proton radiography, achieving remarkable results in some parameter regions.

Despite their successful implementations, current algorithms to solve the proton radiography inverse problem need the exact knowledge of the unmodulated beam intensity profile (source profile) as well as the modulated beam intensity profile. Unfortunately, there is no viable non-destructive method to simultaneously capture the proton beam intensity profiles before and after interacting with the plasma. Moreover, the source profiles are different from shot-to-shot by some considerable amount \cite{manuel-12-prorad-source} making it harder to determine the source profile when the measurement is performed. Only the statistical properties of the source profile variability can be determined reliably, by measuring the beam intensity multiple times without any field interaction \cite{manuel-12-prorad-source,nif-prorad-characteristic-15}.

In this paper, we introduce a statistical method in retrieving the integrated field only by using the statistics of the source profiles, without the exact knowledge of the source profile. Besides retrieving the integrated field, the method also provides the probability distribution of the retrieved field allowing one to perform a statistical analysis of the results, such as checking if the measurement suffers from the inverse problem instability \cite{kasim-18-ipi}.


\section{\label{sec:theory}Theory}

\begin{figure}
    \centering
    \includegraphics[scale=0.3]{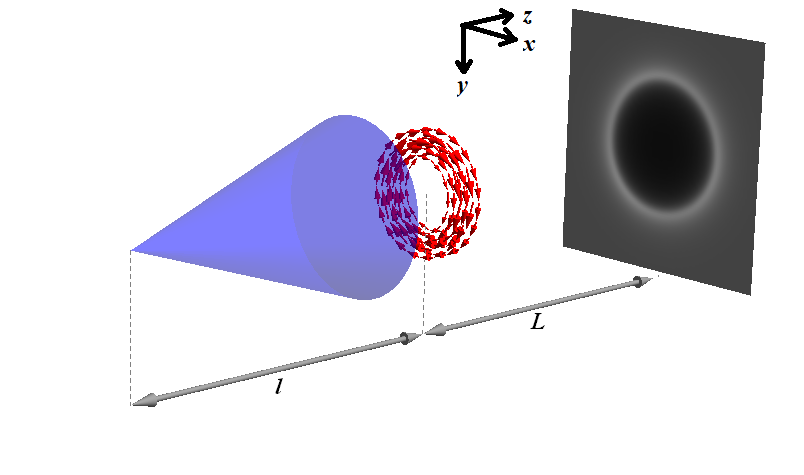}
    \caption{Schematics of a typical set up for proton radiography diagnostics}
    \label{fig:prorad-setup}
\end{figure}

A schematic of a typical proton radiography set-up can be seen in Figure \ref{fig:prorad-setup}. In a typical case, a proton beam is fired from a source through an object with magnetic and/or electric fields. The fields deflect the trajectory of the beam, forming an intensity modulation on the screen behind the object.

Only cases with a single type of dominant field, either magnetic or electric field, is considered in this paper.
It is also assumed that the deflection is small enough so that most of the deflected protons reach the screen.
It is noted that the deflection can still be large enough to reach the non-linear regime.

If a proton beam is fired through a plasma with magnetic field, the beam at coordinate $(x_0, y_0)$ on the object plane will be deflected to an angle
\begin{equation}
    \pmb{\alpha}(x_0, y_0) = -\nabla \Phi(x_0, y_0),
\end{equation}
for $\alpha \ll 1$, where $\Phi$ is the integrated field given by \cite{kugland-12-prorad}
\begin{equation}
    \Phi(x_0, y_0) = -\frac{e}{\sqrt{2mW}} \int \mathbf{A}(x_0, y_0, z_0)\cdot\mathrm{d}\mathbf{z_0}.
\end{equation}
The integrated field, $\Phi$, depends on the charge $e$ and mass $m$ of each proton, the kinetic energy of the beam $W$ and the magnetic vector potential in the object $\mathbf{A}$.
The equation for cases where the electric field is dominant was presented in \cite{kugland-12-prorad}.

The integrated field, $\Phi$, is an important parameter in the retrieval because by knowing $\Phi$ it is straight-forward to retrieve the integrated magnetic field,

\begin{equation}
    \int \mathbf{B}\times\mathrm{d}\mathbf{z_0} = -\frac{\sqrt{2mW}}{e} \nabla \Phi.
\end{equation}

In the cases where the effect from electric field is dominant, the integrated transversal electric field can be retrieved by following the similar steps as above.
For both cases, it is the integrated field profile, $\Phi$, that needs to be retrieved before obtaining the integrated magnetic or electric fields.


If the distance between the object and the screen is $L$, the beam at $(x_0, y_0)$ on the object plane is mapped to the coordinate $(x, y)$ on the screen where they are given by
\begin{align}
\label{eq:x-map}
    x(x_0, y_0) &= x_0 + \pmb{\alpha}(x_0, y_0) \cdot \mathbf{\hat{x}} L, \\
\label{eq:y-map}
    y(x_0, y_0) &= y_0 + \pmb{\alpha}(x_0, y_0) \cdot \mathbf{\hat{y}} L,
\end{align}
assuming the plasma size is much smaller than $L$ and the beam source is collimated. If the beam is diverging from a point source at distance $l$ from the object, then $x_0$ and $y_0$ are replaced by $x_0 \rightarrow x_0 (1 + L/l)$ and $y_0 \rightarrow y_0 (1 + L/l)$ in the equations above.

An analytical equation to determine the intensity on the screen is given by
\begin{equation}\label{eq:intensity-relation}
    I(x, y) = I_0(x,y) \left|\frac{\partial(x,y)}{\partial(x_0, y_0)}\right|^{-1}.
\end{equation}
The term $I_0(x, y)$ denotes the intensity profile without any deflections from the object and the term $|\partial(x,y)/ \partial(x_0, y_0)|$ is the absolute determinant of the Jacobian matrix of $(x,y)$ with respect to $(x_0, y_0)$. If the deflection is large enough, the determinant of the Jacobian can be close to zero at some points and the intensity at the corresponding positions can reach a very high value. These are known as \textit{caustics}.

One way to describe the deflection strength in proton radiography is using a dimensionless variable, as introduced by Kugland \textit{et al.} \cite{kugland-12-prorad}, which is defined as
\begin{equation}
    \mu = L\alpha / a,
\end{equation}
with $L$ the distance from the system to the screen, $\alpha$ the magnitude of the deflection angle, and $a$ the size of the field perturbation.

If $\mu \ll 1$, equation \ref{eq:intensity-relation} can be linearized and the integrated field can be obtained by solving Poisson's equation \cite{kugland-12-prorad},
\begin{equation}
    \nabla^2\Phi \approx \frac{2W}{e}\left(\frac{I}{I_0} - 1\right).
\end{equation}
The two integrals in Poisson's equation solver amplify the lower frequency components of $I/I_0$ by a factor proportional to $1/k^2$ with the wavenumber $k$.

When $\mu$ exceeds a certain value, $\mu \geq \mu_c$, caustics form and parts of the beam cross each other. In this region, the relation between $\Phi$ and $I$ is no longer injective \cite{kugland-12-prorad,bott-17-prorad}. This means that there are multiple profiles of $\Phi$ that correspond to the same intensity profile $I$.

If $\mu$ is on the order of unity and less than $\mu_c$, it gets into the so-called non-linear injective regime. In this regime, Poisson's equation is no longer accurate, but the relation between $\Phi$ and $I$ is still injective. The integrated field, $\Phi$, can be obtained by solving the optimal transport problem: find a way in ``transporting'' the protons from the source profile to resemble the modulated profile with total squared displacement as small as possible. Numerous algorithms are available to solve this type of problem, for example \cite{aurenhammer-1998-minkowski,sulman-11-efficient}.

For the rest of the paper, it is assumed that the proton deflection is always in the injective domain where there is no beam crossing. Field reconstruction in the non-injective domain is beyond the scope of this paper. 

\section{\label{sec:method}Method}

Optimal transport based algorithms to retrieve the integrated field from a source profile and a modulated intensity profile are deterministic. Therefore, the probability of getting $\Phi$ from $I$ and $I_0$ can be written as:
\begin{equation} \label{eq:prob-phi-know-i-i0}
    \mathbb{P}(\Phi | I, I_0) = \delta(\Phi - \Phi'(I,I_0)),
\end{equation}
where $\delta(\cdot)$ is the Dirac delta and $\Phi'(I,I_0)$ is the integrated field profile retrieved using the retrieval algorithms given the modulated intensity profile $I$, and the source profile $I_0$.

When we have no exact knowledge of the source profile $I_0$, one thing that we can do is to marginalize it by integrating the term $I_0$ over its probability distribution,
\begin{equation} \label{eq:prob-phi-know-i}
    \mathbb{P}(\Phi | I) = \int \mathbb{P}(\Phi | I, I_0) \mathbb{P}(I_0)\ \mathrm{d}I_0,
\end{equation}
with $\mathbb{P}(\Phi | I, I_0)$ given by equation \ref{eq:prob-phi-know-i-i0}. From the equation above, we can get the probability distribution of the integrated field $\Phi$ by generating multiple source profile samples according to $\mathbb{P}(I_0)$ and apply the retrieval algorithm for every generated profile. In this paper we use the algorithm from \cite{sulman-11-efficient} to retrieve the integrated field from a modulated intensity profile and a source profile.

Now the problem is shifted to determining the probability distribution of the source profile, $\mathbb{P}(I_0)$. Based on \cite{manuel-12-prorad-source}, the proton profile from a DD and D$^3$He fusion source deviates by several percent from the mean profile and the deviations of two nearby points are correlated. The deviation is larger for proton sources generated by target normal sheath acceleration (TNSA) \cite{warwick2017experimental}. Correlated deviations cannot be captured by a standard normal distribution because it assumes uncorrelated deviations. One way to capture correlated deviations is by representing them using a Gaussian Process (GP) \cite{williams1996gaussian, rasmussen-03-gp}:

\begin{equation}
    \mathbb{P}(I_0 | \theta) = \mathcal{GP}(I_0 | \langle I_0 \rangle, \kappa(\cdot, \cdot | \theta)),
\end{equation}
where $\mathbb{P}(I_0 | \theta)$ is the parameterized prior of the source profile with hyperparameters $\theta$, $\langle I_0 \rangle$ is the mean intensity for the source profile, and $\kappa(\cdot, \cdot | \theta)$ is a kernel function that represents the correlation of two points with hyperparameters $\theta$. The source profile's mean intensity should be equal to the modulated profile's mean intensity.

The explicit expression for GP for an $n$-elements multivariate random variable, $\mathbf{y}\in \mathbb{R}^n$, is given by
\begin{equation}
    \mathcal{GP}(\mathbf{y} | \mu, \kappa) = \frac{\exp{\left[\frac{1}{2}(\mathbf{y}-\mu)^T \mathbf{K}^{-1}(\mathbf{y}-\mu)\right]}}{\sqrt{(2\pi)^n \mathrm{det}(\mathbf{K})}}, 
\end{equation}
where each element in the covariance matrix $\mathbf{K}$ is the value of the kernel $\kappa(\cdot,\cdot)$. In this case, the variable $\mathbf{y}$ is the vectorized source profile $I_0$, and $n$ is the number of pixels. The inputs to the kernel are the position of each pixel in $I_0$.

\begin{figure}
    \centering
    \includegraphics[width=\linewidth]{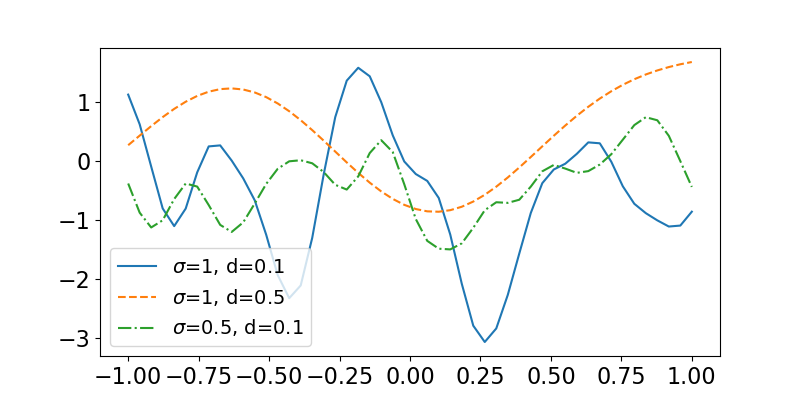}
    \caption{Some samples taken from the one-dimensional Gaussian Process with various values of $\sigma$ and $d$.}
    \label{fig:gp-samples}
\end{figure}

One of the most commonly used kernels is the squared exponential kernel,
\begin{align}
    \kappa_{\mathrm{SE}}(\mathbf{x_1},\mathbf{x_2} | \theta) &= \sigma \exp\left(-\frac{|\mathbf{x_1} - \mathbf{x_2}|^2}{2d^2}\right),
\end{align}
where $\theta = \{\sigma, d\}$ are the hyperparameters of the kernel, $\sigma$ is the expected deviation, and $d$ is the correlated distance. Increasing $\sigma$ will increase the standard deviation from the mean of the profiles. Increasing $d$ will make the deviation less oscillatory. To illustrate the effect of $\sigma$ and $d$, figure \ref{fig:gp-samples} shows some samples taken from the Gaussian Process distribution with various values of $\sigma$ and $d$.

If source profiles are available experimentally, then one can fit the hyperparameters $\theta$ to the experimental source profiles by finding the maximum log-likelihood,
\begin{equation}\label{eq:max-likelihood}
    \theta_{ML} = \arg\max_{\theta'}\sum_{j} \log \mathbb{P}(\hat{I}_{0j} | \theta'),
\end{equation}
where $\hat{I}_{0j}$ is the $j$-th source profile obtained experimentally. However, if there is no experimental data available, one should make a reasonable assumption over the prior probability of the hyperparameters, $\mathbb{P}(\theta)$. This changes Eq. (\ref{eq:prob-phi-know-i}) to
\begin{equation} \label{eq:prob-phi-unknown}
    \mathbb{P}(\Phi | I) = \int \mathbb{P}(\Phi | I, I_0) \mathbb{P}(I_0 | \theta) \mathbb{P}(\theta) \ \mathrm{d}\theta\ \mathrm{d}I_0.
\end{equation}

One way to choose the prior distribution of the hyperparameters when there is no experimental data available is to set it based on known references, for example \cite{manuel-12-prorad-source,nif-prorad-characteristic-15}. Another way is to choose a weak prior of the hyperparameters such as Jeffreys prior \cite{paulo2005gp-prior}, a uniform, or a log uniform prior for certain range.

We present a summary of our method in algorithm \ref{alg:method-summary}.

\begin{algorithm}[H]
    \caption{Retrieving $\Phi$ without explicit $I_0$}
    \label{alg:method-summary}
    \hspace*{\algorithmicindent} \textbf{Input:} the modulated intensity profile $I$ \\
    \hspace*{\algorithmicindent} \textbf{Output:} pool of $\Phi$ samples
    \begin{algorithmic}[1]
        \IF{experimental source profiles available}
            \STATE ~~~ find $\theta_{ML}$ according to equation \ref{eq:max-likelihood}
        \ELSE
            \STATE ~~~ set a prior distribution on $\theta$, $\mathbb{P}(\theta)$
        \ENDIF
        \WHILE{not enough samples}
            \STATE ~~~ use $\theta_s = \theta_{ML}$ or draw a sample for $\theta_s \sim \mathbb{P}(\theta)$
            \STATE ~~~ draw a sample of source profile, $I_{0s} \sim \mathbb{P}(I_0 | \theta_s)$
            \STATE ~~~ retrieve $\Phi_s$ using $I$ and $I_{0s}$
            \STATE ~~~ add $\Phi_s$ to the pool of samples
        \ENDWHILE
    \end{algorithmic}
\end{algorithm}

\begin{figure*}
    \centering
    \includegraphics[width=\textwidth]{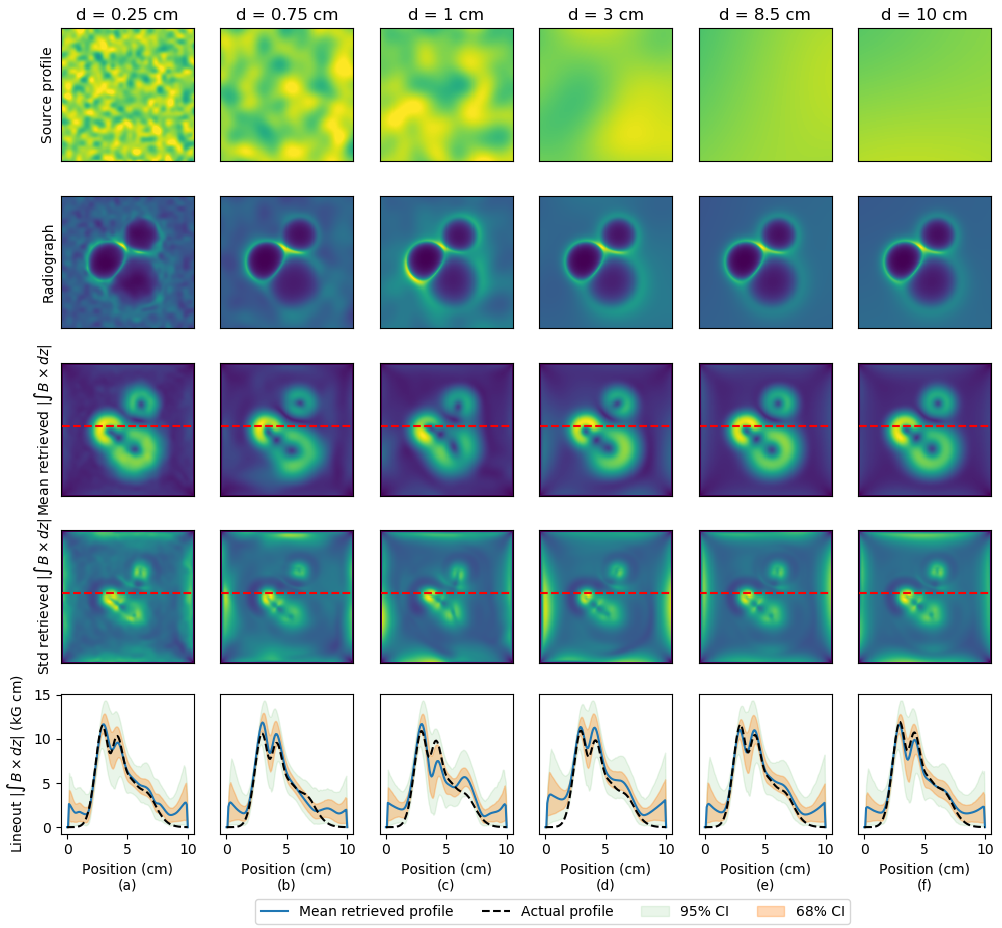}
    \caption{The retrieval results on simulated proton radiography. The simulated modulated intensity profiles (radiographs) are generated from the same integrated magnetic fields and various source profiles with different correlation lengths. The correlation lengths are (a) 0.25 cm, (b) 0.75 cm, (c) 1 cm, (d) 3 cm, (e) 8.5 cm, and (f) 12.5 cm on a $(10\times10)\ \mathrm{cm}^2$ screen. The first row shows the actual source profile used in the simulation (not known by the algorithm). The second row is the radiographs given to the algorithm. The third and forth rows are the average integrated magnetic field profiles and the standard deviations. The red dashed lines on these rows indicate where the lineouts are taken. The last line shows the lineouts of the retrieved profiles compared to the actual profiles (black dashed lines). The orange and green shaded regions on the last row respectively show 68\% ($1\sigma$) and 95\% ($2\sigma$) confidence intervals (CI).}
    \label{fig:synthetic-results}
\end{figure*}

\section{Results}

\subsection{Simulations}

The first test on the proposed method is to retrieve integrated magnetic field statistics from modulated intensity profiles generated from various source profiles and a single integrated magnetic field profile. The source profiles in the test cases have variation about 10\% from the mean value to follow the findings in \cite{nif-prorad-characteristic-15} and have various correlation lengths.
The proton radiography set up follows the experimental set up in \cite{tzeferacos-18}. The distance between the source and the object is 1 cm, object-to-screen distance is 27 cm, and the modulated intensity profile of 15 MeV proton is captured on a screen with size $(10\times10)\ \mathrm{cm}^2$. The magnetic field profile is chosen to give non-linear deflections on the proton beam.

In retrieving the magnetic field in the simulations, we only use the modulated intensity profiles without using the source profiles. Because the actual source profiles are not known in the retrieval, we need to make a prior assumption about the correlation length. One good prior assumption of the correlation length is log-uniform distribution from 10\% to 100\% of the screen size. The range was chosen to capture the variation of source profile that could give considerably different integrated field profiles. It is to obtain an upper estimate of the variance and to avoid over-confidence. The log-uniform distribution is to make it more focused on the shorter correlation length where the largest variation of the retrieved integrated field profiles can be found.

The retrieval results for various source profiles are shown in Figure \ref{fig:synthetic-results}. From the bottom row in the figure, we can see that although in some cases the average retrieved profiles are quite different from the actual profile, the difference can be captured well inside the 95\% confidence interval, except near the boundaries. This shows that the method can capture the uncertainty of the retrieved profiles relatively well. Worse results on the boundaries are because the inversion algorithm \cite{sulman-11-efficient} forces the boundaries to have constant values which reduces the retrieval accuracy near the boundaries.

The mean retrieved profiles are very close to the actual integrated magnetic field profiles when the correlation length is either very short or very long. From the figure, we can say that the average retrieved profiles is relatively far from the actual profile when the correlation length is about 10\% of the screen size (i.e. Figure \ref{fig:synthetic-results}(c)). When the correlation length is increased from 10\%, the average retrieved profiles are getting closer to the actual profile and getting really close when it is about 80\% of the screen size or more. This is because when the correlation length is very short, the source profile is similar to white noise which can be handled very well by the inversion algorithm. When the correlation length is very large, it becomes similar to a uniform source profile.

\subsection{Real experimental data}

\begin{figure}
    \centering
    \includegraphics[width=\linewidth]{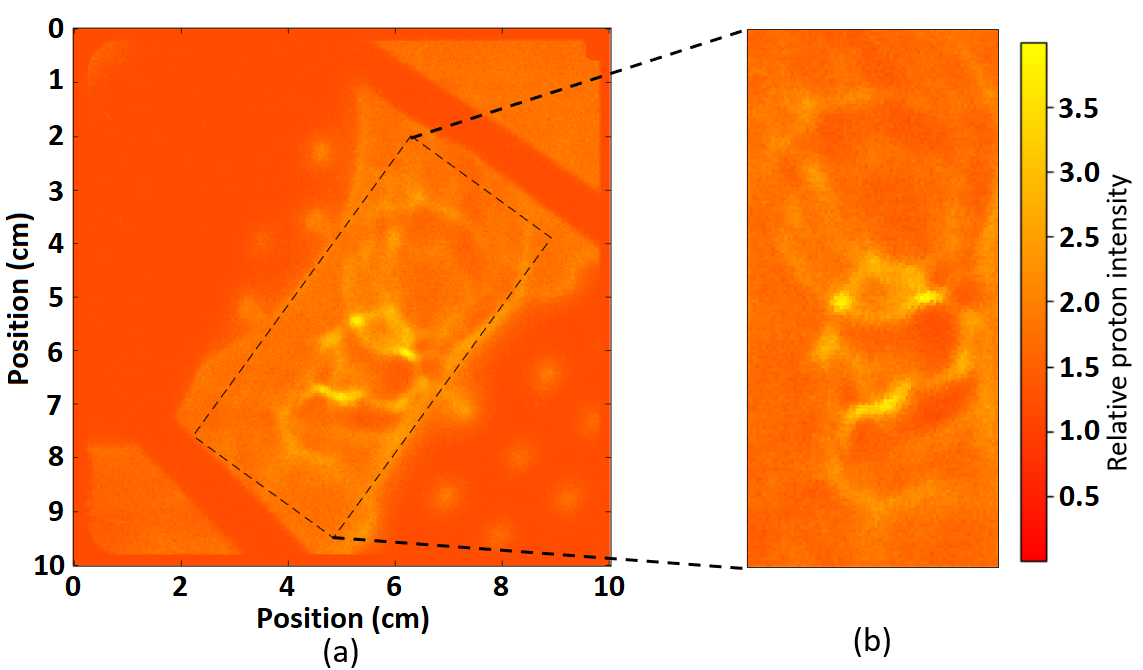}
    \caption{(a) The proton radiography modulated intensity profile captured on an RCF (false color) and (b) the region-of-interest where the proton beam has no obstruction. The data was taken from \cite{tzeferacos-18}.}
    \label{fig:experiment-data}
\end{figure}

\begin{figure*}
    \centering
    \includegraphics[width=\textwidth]{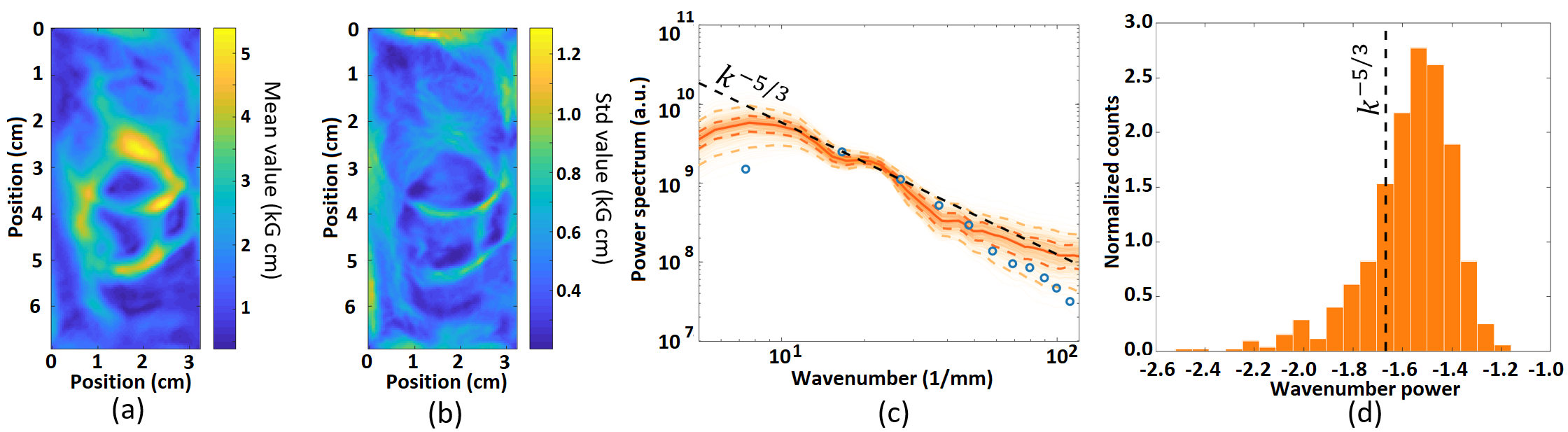}
    \caption{(a) Mean and (b) standard deviation of the retrieved integrated magnetic field from the experiment. (c) The power spectrum of the magnetic field with respect to the wavenumber with blue dots are the data taken from \cite{tzeferacos-18}. The orange solid line indicates the mean values and the dashed lines show the $1\sigma$ and $2\sigma$ intervals. (d) The histogram of the power of the decaying spectrum at large wavenumber.}
    \label{fig:results}
\end{figure*}

We apply our method in retrieving the integrated field as well as the uncertainty to an experiment of dynamo amplification of magnetic fields \cite{tzeferacos-18}. In the experiment, the authors used proton radiography to retrieve the integrated magnetic field using a proton beam generated from a D$^3$He fusion capsule. One of the measured beam intensity profiles is shown in figure \ref{fig:experiment-data}.

Because there is no information on the source profile statistics from this experiment, we take $\sigma$ to be $\sim$10\%, close to the $\sim$13\% standard deviation for proton beams generated from a D$^3$He fusion capsule given in \cite{nif-prorad-characteristic-15}. For the correlated length $d$, we use a prior distribution which is log-uniform from 0.7 cm to 7 cm as in the previous subsection.

In each iteration, we choose a value of $d$ from the prior distribution above, generate a source profile from the Gaussian Process distribution using the chosen hyperparameters, retrieve the integrated field profile using the generated source, and put the integrated field profile into the pool of samples. Figure \ref{fig:results}(a,b) shows the mean and standard deviation of the integrated field from 15,000 samples. From the figure, we can see that with $\sigma = 10\%$, the integrated magnetic field has standard deviation of about $\sim$20\%. The error magnification factor of only 2 is similar to the error propagated by a quadratic equation.

Besides calculating the mean and standard deviation of the integrated field, we also calculated the power spectra from the integrated field samples. This is shown in figure \ref{fig:results}(c), along with the Kolmogorov's power law ($S(k)\propto k^{-5/3}$) for reference. From the figure, we can see that the power spectrum uncertainty is high at small and large wavenumbers. High uncertainty at small wavenumbers is due to the nature of the retrieval algorithm that amplifies low wavenumber elements (i.e. amplification $\sim 1/k^2$). High uncertainty at large wavenumbers is due to limited precision in retrieving features with small amplitudes which happen to be at large wavenumber.

Figure \ref{fig:results}(d) shows the power of the wavenumber in relation to the spectrum. Here we can see that the sample distribution peaks at $\sim k^{-1.5}$ which is shallower than Kolmogorov's power law. We note that the ``true" spectrum of the turbulent magnetic field in the experiment is expected to be even shallower than $\sim k^{-1.5}$ due to lack of injectivity at small scales that manifests diffusively to steepen the measured power spectrum, as shown in \cite{tzeferacos-18}. One needs to include more uncertainty factors to reach a meaningful conclusion, such as non-injectivity of crossing beams and uncertainty in proton energy, which are not the focus of this paper.

\section{\label{sec:conclusion} Conclusions}
We have presented a statistical method to retrieve integrated fields from proton radiography without knowing the exact source profile. The probability distribution of the integrated field is obtained by marginalizing out the probability distribution of the source profile. The distribution of source profiles can be obtained by collecting a number of proton radiography samples without any interaction with electric and magnetic fields, or by making a weak prior assumption. The method has been applied to an experiment to retrieve the integrated magnetic field and its statistics, showing the robustness of the proposed approach.

\input acknowledgement.tex   
\section*{Code availability}
The code associated with this paper can be found on: https://github.com/OxfordHED/proton-radiography-no-source/

\input bibliography.bbl

\end{document}

%% file: authorlist.tex
\author{M. F. Kasim}
\email{muhammad.kasim@physics.ox.ac.uk}
\author{A. F. A. Bott}
\affiliation{Clarendon Laboratory, Department of Physics, University of Oxford, Parks Road, Oxford OX1 3PU, United Kingdom}
\author{P. Tzeferacos}
\author{D. Q. Lamb}
\affiliation{Department of Astronomy and Astrophysics, University of Chicago, 5640S. Ellis Ave, Chicago, IL 60637, USA}
\author{G. Gregori}
\author{S. M. Vinko}
\affiliation{Clarendon Laboratory, Department of Physics, University of Oxford, Parks Road, Oxford OX1 3PU, United Kingdom}

%% file: acknowledgement.tex
%
\section*{Acknowledgement}
S.M.V. is grateful for support from the Royal Society. M.F.K. and S.M.V. acknowledge support from the UK EPSRC grant EP/P015794/1. G.G. and A.F.A.B. acknowledges support from AWE plc., and the UK EPSRC (EP/M022331/1 and EP/N014472/1). This work was supported in part from the U.S. DOE under Cooperative Agreement DE-NA0001944 to the University of Rochester, Field Work Proposal 57789 to Argonne National Laboratory, and grants DE-NA0002724, DE-NA0003605, and DE-SC0016566 to the University of Chicago; and from the National Science Foundation under grant PHY-1619573. We acknowledge support from DOE NNSA under subcontracts B632670 and 536203 with the University of Chicago.